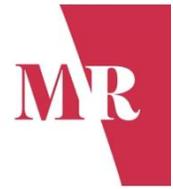

*Commentary*

# A dual typology of social media interventions and deterrence mechanisms against misinformation

*In response to the escalating threat of misinformation, social media platforms have introduced a wide range of interventions aimed at reducing the spread and influence of false information. However, there is a lack of a coherent macro-level perspective that explains how these interventions operate independently and collectively. To address this gap, I offer a dual typology through a spectrum of interventions aligned with deterrence theory and drawing parallels from international relations, military, cybersecurity, and public health. I argue that five major types of platform interventions, including removal, reduction, informing, composite, and multimodal, can be mapped to five corresponding deterrence mechanisms—hard, situational, soft, integrated, and mixed deterrence—based on purpose and perceptibility. These mappings illuminate how platforms apply varying degrees of deterrence mechanisms to influence user behavior.*

Authors: Amir Karami (1)
Affiliations: (1) School of Data Science and Analytics, Kennesaw State University, USA

## Introduction

In response to the growing threat of misinformation, social media platforms have deployed diverse interventions designed to mitigate the spread and influence of false or misleading information (Krishnan et al., 2021) and have reached nearly half of all users in the process (Saltz et al., 2021). These interventions—such as content removal, limiting the visibility of malicious activities, and attaching warning labels to posts—reflect a dynamic governance landscape shaped by public pressure, political scrutiny, and evolving platform capabilities (Ng et al., 2021; Zannettou, 2021).

Despite the increasing number and variety of interventions, there is a lack of a coherent macro-level typology that explains how these interventions operate independently and collectively. Current studies tend to examine platform interventions in isolation or emphasize their technical aspects (Broniatowski et al. 2023; Vincent et al., 2022; Pennycook & Rand, 2019), rather than analyzing their intended impact or positioning them within broader deterrence or governance models. A macro-level perspective is necessary for understanding how platforms seek to shape user behavior at scale.

---

[1] *A publication of the Shorenstein Center on Media, Politics and Public Policy at Harvard University, John F. Kennedy School of Government.*

To address this gap, I propose a dual typology that systematically links social media interventions to deterrence mechanisms. This approach considers two core dimensions of interventions: purpose (coercion, restriction, persuasion, or synthesis) and perceptibility (visibility to users). The framework aims to clarify how interventions function not only as technical solutions but also as behavioral strategies that deter misinformation at scale.

## Deterrence theory foundation

Deterrence theory in criminology posits that individuals can be dissuaded from unwanted behavior if potential punishments are sufficiently certain, swift, and severe (Tomlinson, 2016). Over time, this foundational concept has evolved into a broader typology of deterrence strategies, each defined by the mechanisms it employs. In international relations, hard deterrence (power) refers to coercive behavior through force or penalties, while soft deterrence relies on persuasive strategies by appealing to users' values, beliefs, or understanding of consequences (Nye, 2005). Lying between these two approaches is situational deterrence (Cusson, 1993), which focuses on limiting opportunities for undesirable behavior without resorting to coercion. Expanding beyond single tactics, integrated deterrence frameworks in military and cybersecurity domains combine multiple capabilities into a single, cohesive strategy of deterrence (Chen, 2023; Stewart, 2024). Likewise, mixed deterrence, frequently seen in space and defense applications, refers to the simultaneous application of different deterrent mechanisms to address complex or hybrid threats (Arkin, 1986; Chen, 2023). Integrated deterrence combines multiple tactics within one intervention on the same entity, while mixed deterrence applies different mechanisms across entities or levels to build a broader defense. Table 1 offers a definition for each deterrence mechanism.

*Table 1. Deterrence mechanisms.*

| Term | Definition |
| --- | --- |
| Hard deterrence | Coercing behavior through force or penalties (Nye, 2005) |
| Soft deterrence | Persuading behavior by appealing to users' values, beliefs, or understanding of consequences (Nye, 2005) |
| Situational deterrence | Restricting opportunities for undesirable behavior without resorting to coercion (Cusson, 1993) |
| Integrated deterrence | Combining multiple capabilities into a single, cohesive deterrence strategy (Chen, 2023; Stewart, 2024) |
| Mixed deterrence | Adopting simultaneous application of different deterrent mechanisms to address complex or hybrid threats (Arkin, 1986; Chen, 2023) |

## Typology of social media interventions and deterrence mechanisms

I propose a dual typology (see Table 3) that maps intervention types with corresponding deterrence mechanisms, offering a more systematic understanding of how social media platforms combat misinformation. This bridges the technical implementation and behavioral function of interventions, highlighting how each type of intervention serves a different purpose (see Table 2) and deterrence mechanism (see Table 1). The typology is grounded in two dimensions:

- *Purpose*: the strategic intent—whether the intervention seeks to coerce, restrict, or persuade, or synthesize these three purposes.
- *Perceptibility*: the degree to which the intervention is visible or noticeable to users. In the typology, the high and low perceptibility reflect broad patterns in how interventions are typically experienced by users. For example, informing interventions such as warning labels are usually highly perceptible because they appear directly on the content that users view, while reduction interventions such as downranking are less perceptible because they operate algorithmically in the background. At the same time, perceptibility is context-dependent. The same intervention may be more or less visible depending on the level at which it is applied. For instance, the removal of a single post may be relatively invisible to the wider community, whereas the suspension of an entire account is highly noticeable to both the affected user and their audience.

*Table 2*. The definition of social media interventions for fighting misinformation.

| Term | Definition |
| --- | --- |
| Removal intervention | Deletion of content or the suspension/removal of user accounts (Center for an Informed Public et al., 2021) |
| Informing intervention | Providing users with information, context, or warnings about content (Center for an Informed Public et al., 2021) |
| Reduction intervention | Curtailing the reach or visibility of content and accounts (Center for an Informed Public et al., 2021) |
| Composite intervention | Combining multiple types of interventions into a single, unified approach that is applied together as one strategy (Glasziou et al., 2014) |
| Multimodal intervention | Adopting two or more interventions across different levels of action in a coordinated way (Burgener et al., 2008; Morton et al., 2020) |

*Table 3*. Dual typology of deterrence mechanisms and social media interventions.

| Deterrence mechanism | Intervention | Purpose | Perceptibility | Example |
| --- | --- | --- | --- | --- |
| Hard | Removal | Coercive | High | Account suspension |
| Soft | Informing | Persuasive | High | Warning |
| Situational | Reduction | Restrictive | Low | Downranking |
| Integrated | Composite | Restrictive + Persuasive | High | Labeling content and disabling its engagement features |
| Mixed | Multimodal | Coercive/Restrictive/Persuasive | High | Simultaneous downranking some comments and removing other comments |

*Hard deterrence mechanism via removal intervention*

Building on this framework, I now illustrate each intervention type through concrete examples, beginning with the most restrictive and perceptible form of moderation. Removal refers to deleting content or suspending and removing user accounts or online communities that disseminate misinformation (Center for an Informed Public et al., 2021; Cima et al., 2024). This represents the most restrictive form of intervention, as it fully blocks access to content, the offending user, or the community sharing misinformation. All major platforms employ removal for content that violates policies. For instance, YouTube, Facebook, and Twitter (now X) removed the "Plandemic" video—a viral COVID-19 conspiracy theory—in May 2020 to staunch its spread (Culliford, 2020a). Entire pages or accounts dedicated to spreading falsehoods have been taken down as well. Removal is a hard deterrence mechanism because it operates through direct coercion with high perceptibility: it imposes a high cost on the violator by erasing their content or presence, thereby unequivocally signaling that the behavior is prohibited.

*Soft deterrence mechanism via informing intervention*

*Informing interventions* aim to educate users or provide additional context without removing content. These include warning labels, fact-check notices, banners, contextual panels, and interstitial pop-ups that appear before viewing or sharing flagged posts, along with other messages designed to educate or caution users (Center for an Informed Public et al., 2021). For example, Facebook and Instagram apply labels to posts debunked by fact-checkers, often covering the post with a warning that must be clicked through. Twitter has used interstitial warnings on tweets and prompts (Geeng et al., 2020). TikTok has added banners on COVID-19 or vaccine-related posts with reminders and links to authoritative information (Morgan, 2020). These measures allow the content to remain accessible (no deletion or reach reduction in many cases), thus fully preserving the user's ability to speak and others' ability to hear them, but they inject additional information to steer perception. Informing is the epitome of a soft deterrence mechanism: it relies on persuasion rather than any material restriction.

*Situational deterrence mechanism via reduction intervention*

*Reduction* refers to interventions that curtail the reach or visibility of content and accounts associated with misinformation, without fully removing them (Center for an Informed Public et al., 2021). Tactics include downranking or demoting posts in algorithmic feeds, limiting the distribution of certain URLs or stories, or placing frictions on sharing (Center for an Informed Public et al., 2021). The content remains on the platform, but it is harder to encounter. For example, Facebook's feed algorithm downranks posts identified as containing "exaggerated or sensational health claims" so that they appear to far fewer users (Yeh, 2019). Twitter has in the past downranked replies or tweets deemed misleading (Roth & Harvey, 2018; Roth & Pickles, 2020). Reduction is a situational deterrence mechanism: it imposes restrictions on the opportunity of misinformation spread but not an absolute ban. In terms of coerciveness, it is less coercive than hard deterrence yet more coercive than soft deterrence. Notably, reduction measures are low perceptibility.

*Integrated deterrence mechanism via composite intervention*

*Composite intervention*, in this taxonomy, refers to combining multiple types of interventions (Glasziou et al., 2014), spanning soft and situational deterrence mechanisms, into a single coordinated strategy. Rather than a single mode of intervention, an integrated deterrence mechanism synthesizes tactics to address misinformation more holistically. In practical terms, this could mean applying two or more interventions simultaneously to the same content or actor. For example, Twitter's handling of certain election misinformation in 2020 was integrated: it placed warning labels on tweets and disabled the ability to like or retweet those tweets without quote-commenting (Culliford, 2020b). Reddit's "quarantine" feature for problematic communities is another integrated intervention: a quarantined subreddit is not banned, but it is placed behind a click-through warning page and is kept out of search results and recommendations (Reddit, 2018). In short, composite intervention is an integrated deterrence mechanism because it deliberately combines two deterrence modes to create a more effective or targeted overall deterrent. The combination of interventions can thus be seen as creating a composite intervention that attacks the misinformation's spread through both restriction and persuasion.

*Mixed deterrence mechanism via multimodal intervention*

*Multimodal intervention* refers to strategies that strengthen the overall information environment or user resilience against misinformation through multiple interventions. In the typology, this aligns with mixed deterrence mechanisms, a comprehensive approach where multiple layers of defense and influence are deployed in a coordinated way. Unlike integrated intervention, which bundles two actions on a single piece of content, mixed deterrence mechanisms span across different levels (post, account, community, network, and off-platform), creating a repetitive or structural reinforcement. For example, Twitter began removal interventions in 2017, in which violating posts were deleted at the content level. In 2018, it introduced reduction interventions that downranked or limited the visibility of misleading content. By 2020, Twitter also adopted informing interventions, such as applying warning labels or prompts to provide additional context (Crowell, 2017; Culliford, 2020b; Roth & Harvey, 2018). Overall, a mixed deterrence mechanism means deploying multiple interventions that work together to form a robust deterrent and mitigation structure, thereby leveraging varying degrees of coerciveness, restrictiveness, and persuasiveness. This approach underscores the rationale for organizing interventions according to their underlying deterrence mechanism. Unlike other interventions, which typically target specific entities such as a post, a user, or a community, multimodal interventions operate across multiple levels of the platform. They may simultaneously combine actions at the post, account, and community levels, creating a layered strategy that reinforces deterrence across the broader information environment. This distinction frames multimodal interventions as a comprehensive moderation approach that operates across multiple layers of the platform, rather than a tactic aimed at a single entity.

# Conclusion

Organizing social media misinformation interventions into five categories (removal, reduction, informing, composite, and multimodal) and mapping these categories onto five deterrence mechanisms (hard, situational, soft, integrated, and mixed) provides a theoretically grounded framework that clarifies how each strategy functions to influence user behavior. This typology highlights two core dimensions: purpose (coercive, restrictive, persuasive, or blended) and perceptibility (how visible the intervention is to users). Classical deterrence theory serves as the foundational logic of this framework, offering structured insight into the behavioral assumptions behind various intervention types. Social media governance research emphasizes that interventions are not "one size fits all" in either their effectiveness or public reception. This reinforces the need to distinguish interventions not only by their technical function but also by their social acceptability, level of intrusiveness, and communicative transparency. The typology also acknowledges the importance of combining hard and soft tactics through integrated or mixed deterrence approaches. These strategies may be especially effective when confronting hybrid threats such as politicized misinformation or coordinated influence campaigns.

Beyond its theoretical value, this typology has practical relevance for a range of stakeholders. Researchers can use it as a coding framework to systematically track different deterrence mechanisms across platforms. Journalists and fact-checkers can draw on it to explain these mechanisms in accessible terms for the public. Policy analysts and regulators may apply it to assess whether measures such as downranking misleading political ads strike the right balance between limiting harm and preserving open debate. Platform practitioners can use the typology to guide the design of interventions, while civil society organizations can leverage it to advocate for layered approaches that combine removal, reduction, and informing to counter coordinated influence campaigns. These examples illustrate the flexibility of the typology, since different deterrence mechanisms and interventions can be applied depending on the context and stakeholder goals. In short, by mapping social media interventions onto the dual typology, I gain not only a coherent and flexible framework for scholarly analysis but also a practical lens for evaluating, communicating, and refining responses to misinformation across health, political, and crisis domains.

# Bibliography

Arkin, W. M. (1986). The new mix of defense and deterrence. *Bulletin of the Atomic Scientists*, *42*(6), 4–5. https://doi.org/10.1080/00963402.1986.11459379

Broniatowski, D. A., Simons, J. R., Gu, J., Jamison, A. M., & Abroms, L. C. (2023). The efficacy of Facebook's vaccine misinformation policies and architecture during the COVID-19 pandemic. *Science Advances*, *9*(37), Article eadh2132. https://doi.org/10.1126/sciadv.adh2132

Burgener, S. C., Yang, Y., Gilbert, R., & Marsh-Yant, S. (2008). The effects of a multimodal intervention on outcomes of persons with early-stage dementia. *American Journal of Alzheimer's Disease & Other Dementias®*, *23*(4), 382–394. https://doi.org/10.1177/1533317508317527

Center for an Informed Public, Digital Forensic Research Lab, Graphika, & Stanford Internet Observatory. (2021). *The long fuse: Misinformation and the 2020 election*. Stanford Digital Repository: Election Integrity Partnership. https://purl.stanford.edu/tr171zs0069

Chen, J. Q. (2023). Deterrence in cyberspace: An essential component in integrated deterrence. In J. L. Billingsly (Ed.), *Integrated deterrence and cyberspace: Selected essays exploring the role of cyber operations in the pursuit of national interest* (pp. 1–22). National Defense University Press. https://digitalcommons.ndu.edu/cgi/viewcontent.cgi?article=1001&context=strategic-monographs

Crowell, C. (2017, June 14). *Our approach to bots and misinformation*. X. https://blog.x.com/en_us/topics/company/2017/Our-Approach-Bots-Misinformation

Culliford, E. (2020a, May 7). *Facebook, YouTube remove "Plandemclaimsic" video with "unsubstantiated" coronavirus claims.* Reuters. https://www.reuters.com/article/technology/facebook-youtube-remove-plandemic-video-with-unsubstantiated-coronavirus-cl-idUSKBN22K073/

Culliford, E. (2020b, October 9). *Twitter imposes restrictions, more warning labels ahead of U.S. election*. Reuters. https://www.reuters.com/article/world/twitter-imposes-restrictions-more-warning-labels-ahead-of-us-election-idUSKBN26U1X9/

Geeng, C., Francisco, T., West, J., & Roesner, F. (2020). *Social media COVID-19 misinformation interventions viewed positively, but have limited impact*. arXiv. https://doi.org/10.48550/arXiv.2012.11055

Glasziou, P. P., Chalmers, I., Green, S., & Michie, S. (2014). Intervention synthesis: A missing link between a systematic review and practical treatment(s). *PLoS Medicine*, *11*(8), Article e1001690. https://doi.org/10.1371/journal.pmed.1001690

Krishnan, N., Gu, J., Tromble, R., & Abroms, L. C. (2021). Research note: Examining how various social media platforms have responded to COVID-19 misinformation. *Harvard Kennedy School (HKS) Misinformation Review*, *2*(6). https://doi.org/10.37016/mr-2020-85

Morgan, K. (2020, December 15). *Taking action against COVID-19 vaccine misinformation*. TikTok. https://newsroom.tiktok.com/taking-action-against-covid-19-vaccine-misinformation?utm_source=chatgpt.com&lang=en-GB

Karami 7Morton, D. P., Hinze, J., Craig, B., Herman, W., Kent, L., Beamish, P., Renfrew, M., & Przybylko, G. (2020). A multimodal intervention for improving the mental health and emotional well-being of college students. *American Journal of Lifestyle Medicine*, *14*(2), 216–224. https://doi.org/10.1177/1559827617733941

Ng, K. C., Tang, J., & Lee, D. (2021). The effect of platform intervention policies on fake news dissemination and survival: An empirical examination. *Journal of Management Information Systems*, *38*(4), 898–930. https://doi.org/10.1080/07421222.2021.1990612

Nye, J. S. (2005). *Soft power: The means to success in world politics*. Public Affairs Books. https://www.wcfia.harvard.edu/publications/soft-power-means-success-world-politics

Reddit. (2018). *Revamping the quarantine function* [Online form post]. Reddit. https://www.reddit.com/r/announcements/comments/9jf8nh/revamping_the_quarantine_function/

Roth, Y., & Harvey, D. (2018). *How Twitter is fighting spam and malicious automation.* X. https://blog.x.com/en_us/topics/company/2018/how-twitter-is-fighting-spam-and-malicious-automation

Roth, Y., & Pickles, N. (2020). *Updating our approach to misleading information*. X. https://blog.x.com/en_us/topics/product/2020/updating-our-approach-to-misleading-information

Pennycook, G., & Rand, D. G. (2019). Fighting misinformation on social media using crowdsourced judgments of news source quality. *Proceedings of the National Academy of Sciences*, *116*(7), 2521–2526. https://doi.org/10.1073/pnas.1806781116

Saltz, E., Barari, S., Leibowicz, C., & Wardle, C. (2021). Misinformation interventions are common, divisive, and poorly understood. *Harvard Kennedy School (HKS) Misinformation Review*, *2*(5). https://doi.org/10.37016/mr-2020-81

Stewart, C. (2024). *Think bigger, act larger: A US-Australia led coalition for a combined joint deterrence force in the Indo-Pacific*. CEIP: Carnegie Endowment for International Peace. https://coilink.org/20.500.12592/4xahwi9

Tomlinson, K. D. (2016). An examination of deterrence theory: Where do we stand? *Federal Probation*, *80*, 33–38. https://www.uscourts.gov/sites/default/files/80_3_4_0.pdf

Vincent, E. M., Théro, H., & Shabayek, S. (2022). Measuring the effect of Facebook's downranking interventions against groups and websites that repeatedly share misinformation. *Harvard Kennedy School Misinformation Review*. *3*(3). https://doi.org/10.37016/mr-2020-100

Yeh, T. (2019). *Addressing sensational health claims*. Meta Newsroom. https://about.fb.com/news/2019/07/addressing-sensational-health-claims/

Zannettou, S. (2021). "I won the election!": An empirical analysis of soft moderation interventions on Twitter. *Proceedings of the International AAAI Conference on Web and Social Media*, *15*(1), 865–876. https://doi.org/10.1609/icwsm.v15i1.18110

**Funding**
No funding has been received to conduct this research.

**Competing interests**
The author declares no competing interests.